# One number per hour is not enough: directional carbon intensity of data-center ramps on the PJM fleet

Xiaoying Shi[1,2,3], Jinyue Yan[1,2,*]

1 International Centre of Urban Energy Nexus, The Hong Kong Polytechnic University, Kowloon, Hong Kong, China.

2 State Key Laboratory of Climate Resilience for Coastal Cities, Department of Building Environment and Energy Engineering, The Hong Kong Polytechnic University, Hong Kong, China.

3 Lead contact

*Correspondence: jjyan@polyu.edu.hk

Summary

A data-center campus can change its load by hundreds of megawatts within an hour, yet every carbon signal in use gives one value per hour for a load change, whatever its size or direction. We compute the directional carbon intensity, the emission caused or avoided per megawatt-hour of a finite upward or downward ramp, for every hour of 2025 on the PJM Interconnection's fossil fleet, and compare it with the locational marginal emission rate (LME) and the regional average. The LME misstates a 1 GW ramp by more than 10% in 60% to 83% of hours in eleven months. Upward and downward ramps differ in a quarter to three quarters of hours but agree on their monthly mean. Location-based reporting assigns 47% to 59% of a ramp's emission to other consumers. A campus should schedule on the signal for its ramp's size and direction and report the emission it causes.

Graphical abstract

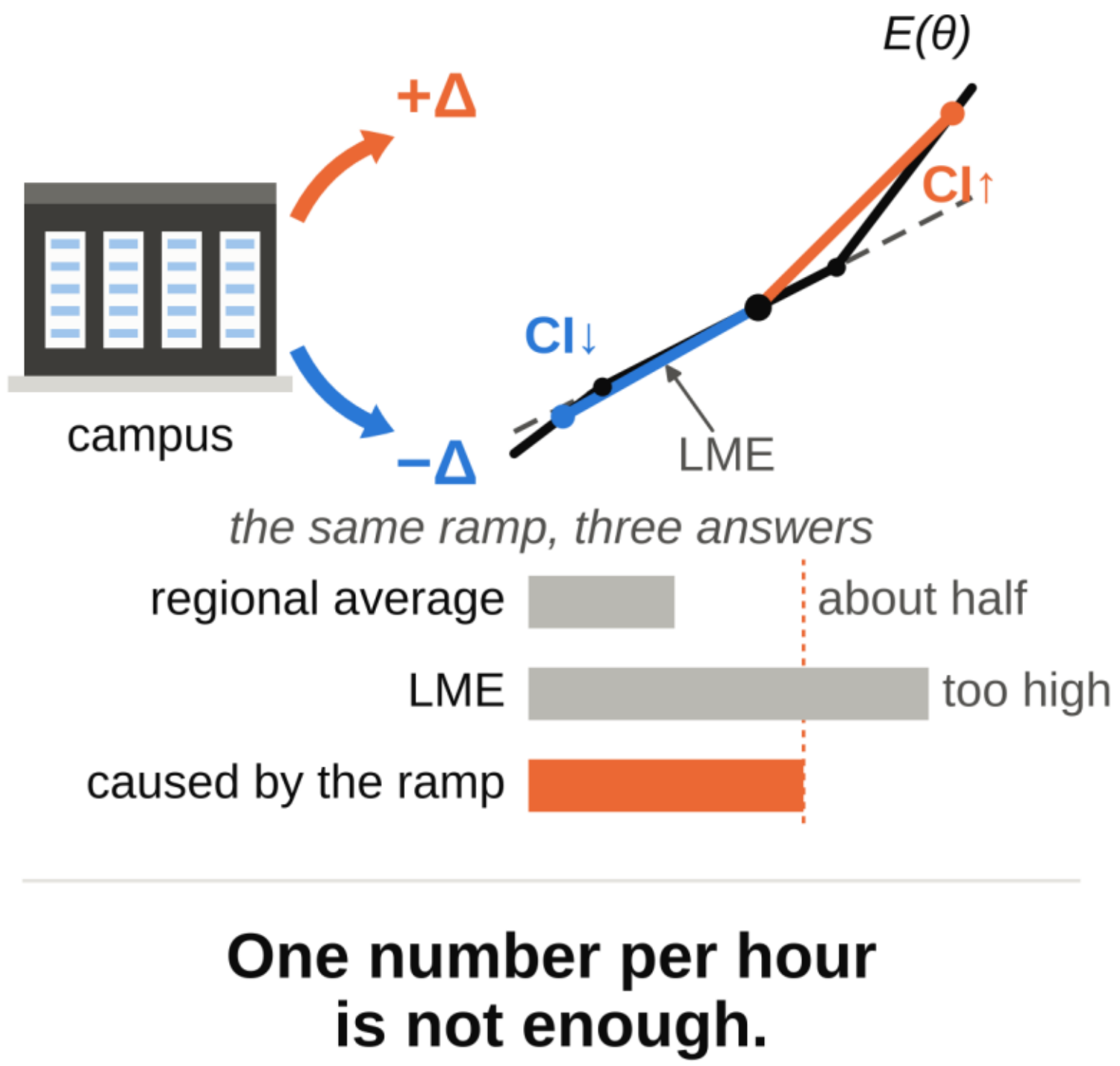

## Context & Scale

Data centers are becoming some of the largest and most flexible electricity consumers on the power grid. When a campus raises or lowers its load by tens to hundreds of megawatts, power plants elsewhere on the grid change their output to balance it, and the carbon dioxide those plants emit or avoid is the climate effect of that decision. Grid operators and regulators measure this effect with carbon signals: one number per hour that says how clean the grid is at that moment. Different signals answer different questions. An average signal divides the emissions the grid produced among all its consumers in proportion to their use and is used for reporting; a marginal signal estimates how emissions would change if demand at one place changed and is used for scheduling. Neither was designed for a load that moves by hundreds of megawatts in one step: a change that large is met by several power plants with different emission rates, not by the single plant at the margin, so one number cannot describe it. Using the real power-plant fleet of the largest US grid operator over a full year, we measure how far each signal strays from the emission a ramp causes, and show that the answer depends on the size of the ramp and on its direction. The result matters to grid operators deciding what signal to publish, to data-center operators scheduling flexible load, and to the bodies now rewriting the rules by which electricity emissions are reported.

Highlights

1. The marginal carbon signal misstates 1 GW data-center ramps in 60 to 83% of hours
2. Up and down ramps differ in a quarter to three quarters of hours but agree on average
3. Location-based reporting assigns about half of a ramp's emission to other consumers
4. PJM's published five-minute carbon signal keeps its state less than half the time

## INTRODUCTION

Electricity consumed by data centers is projected to more than double, from about 415 TWh in 2024 to about 945 TWh in 2030 [1]. Data center campus is generally concentrated in a few grid regions, and the Dominion zone of the PJM Interconnection, which covers most of Virginia, serves the largest concentration of data centers in the world [2]. Unlike most large loads, part of a data center's demand can be shifted in time and between sites, and operators already use this flexibility for carbon goals: Google shifts flexible computing to hours with lower grid carbon intensity [3]. Both the scheduling and the reporting of data centers depend on carbon signals: the former expects a marginal signal to estimate the emission that a load change causes, and the latter uses average grid emission factors to declare Scope 2 emission under the Greenhouse Gas Protocol [4]. Data-center load in a single area can change by gigawatts within seconds: on 22 July 2026 a transmission fault in Northern Virginia moved about 3.8 GW of data-center load to on-site backup supply, the largest such event in PJM's history [5], and NERC has recorded more than 30 unplanned data-center load-loss events since 2023 [6].

Two families of carbon signal answer different questions. Attributional signals, such as the regional average intensity used for location-based Scope 2 reporting [4] and nodal intensities from carbon emission flow [7], divide the emissions that generators produced among consumers. Consequential signals, such as marginal emission factors [8] and locational marginal emissions (LMEs) [9], estimate the change in system emissions caused by a change in demand, and have been used to guide data-center load shifting [10]. The two do not overlap in function and cannot substitute for each other. In a test-system study of data-center load shifting, LME-guided shifting was the only strategy that reduced system emissions, whereas shifting guided by average-based signals left total emissions almost unchanged or slightly higher and moved about 1% of accounted emissions onto other loads; LMEs used for accounting assigned more than twice the emissions the system produced [11]. A hyperscale data-center campus therefore needs both signals: a consequential one to schedule and an attributional one to report.

A consequential signal is a derivative of system emissions with respect to load, so it is exact only while the same generators and constraints remain at the margin. Nilges et al. computed the range of load increase and decrease over which an LME stays exact, and found it mostly about 5 MW in a German transmission model [12]. Khanpour et al. bounded how far emissions under uncertain demand can exceed the value predicted from LMEs, separating the variation while the same constraints remain binding, the probability that a different set of constraints becomes binding, and the error in estimating the LMEs [13]. Cote and Sun added a 200 MW load, sized as a large data center, at a randomly chosen bus and hour of a 1,493-bus western US model, repeated this 1,000 times, and found that the summed emission increase was 114% of the sum predicted by LMEs [14]. A sum can be close even when individual trials are not, because trials in which the LME predicts too much cancel trials in which it predicts too little. A campus schedules hour by hour, so the error that matters is the error in each hour.

To the best of our knowledge, no study has measured the hourly error of the marginal signal for a data-center campus ramp on a real generation fleet, whether this error differs between upward and downward ramps, or how a grid operator's published nodal signal compares with the emission a ramp causes. A validity range shows when the LME stops being exact [12]; a campus also needs to know how wrong it then is. Here we measure this with the directional carbon intensity, defined as the change in system emissions per megawatt of a finite upward or downward ramp at the campus [15]. As the ramp shrinks to zero, the directional intensity converges to the one-sided LME, and within the validity range the two are equal [15]. We compute the directional carbon intensity and the LME for every hour of 2025 on the fossil fleet of the PJM Interconnection: 682 units reported to the U.S. Environmental Protection Agency, with their observed hourly commitment and measured emission rates, dispatched in merit order without transmission limits, for campus ramps of 50 MW to 1 GW in each direction. We then compare both with the five-minute nodal marginal emission rate that PJM publishes for a data-center node in Ashburn, Virginia. This design answers four questions: how far the LME departs from the emission of a ramp as the ramp grows, whether that departure depends on the direction of the ramp, how much of the emission a ramp causes is assigned to the campus under location-based reporting, and how the signal that PJM publishes compares with both.

We find that the LME misstates the emission of a ramp more often as the ramp size grows: by more than 10% in 15% to 21% of hours for a 100 MW ramp depending on the month, and in 60% to 83% of hours for a 1 GW ramp in every month except July (42%), even when every committed unit is free to respond. Upward and downward ramps of the same size differ hour by hour, yet their monthly average emission nearly agree, so a monthly average hides the difference that matters for scheduling. Location-based reporting assigns the campus 47% to 59% less than the emission its ramps cause, depending on the month. PJM's published five-minute signal persists only weakly from one interval to the next. A campus should therefore schedule on a signal stated for the size and direction of its ramp, and report the emission its ramp causes alongside the location-based figure.

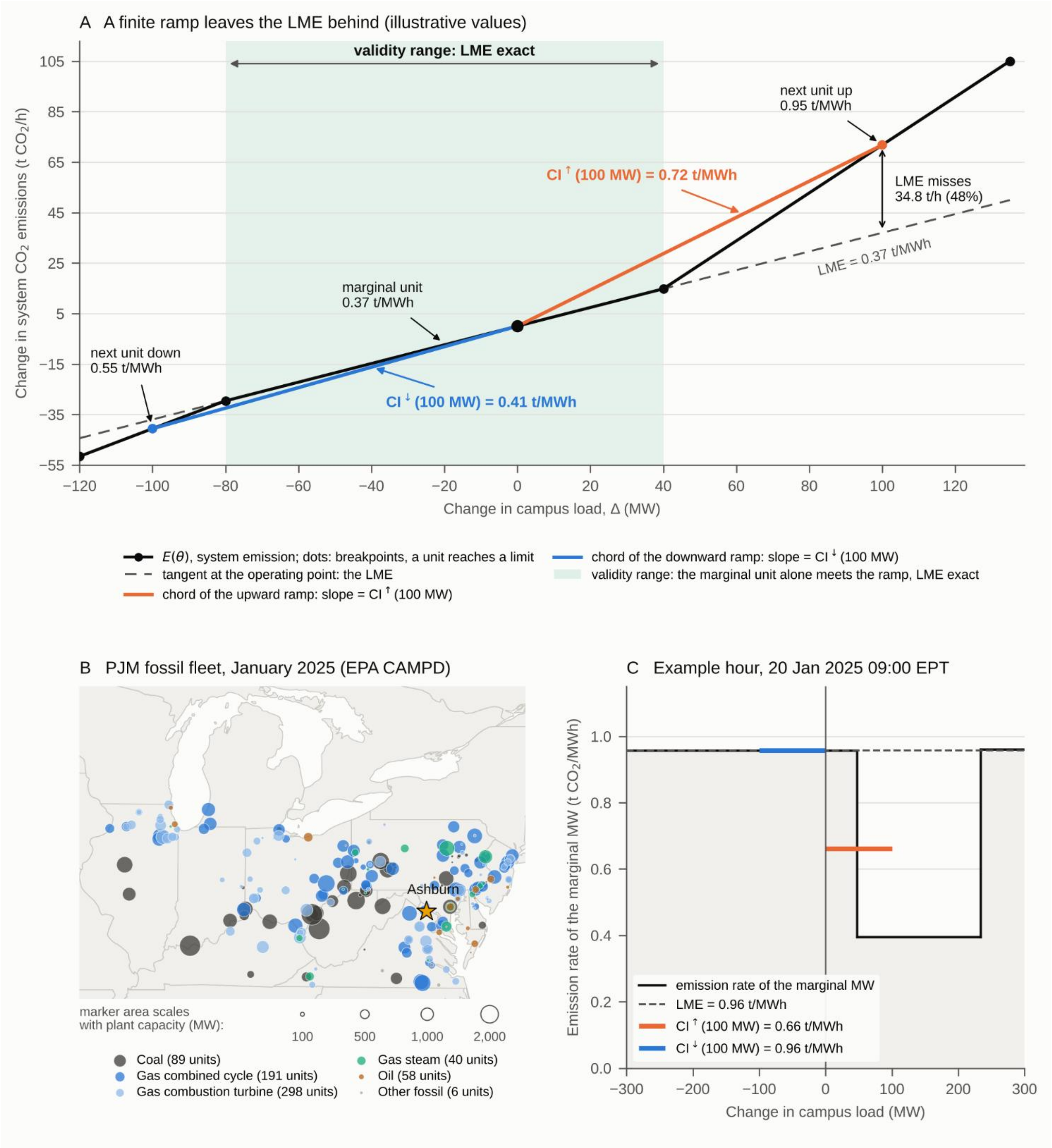

Figure 1. Directional carbon intensity of a data-center ramp: definition, the PJM fleet, and one hour (A) Schematic: the LME is the slope at the operating point; a finite ramp follows the chord (illustrative values). (B) The PJM fossil fleet, January 2025, by fuel and capacity. (C) One hour of the real fleet: the marginal emission rate either side of the operating point.

## RESULTS

### A finite ramp leaves the marginal signal behind

Figure 1 sets out the quantity we measure. The emission of the system dispatch $E_h(\theta)$ is a piecewise linear function of the campus load θ: within each linear piece the same unit is marginal, and at each breakpoint the marginal unit changes. The LME is the slope of the piece that contains the operating point. A finite upward ramp of size Δ causes $E_h(\theta_0+\Delta)-E_h(\theta_0)$, the chord from the operating point to the end of the ramp. As shown in Figure 1A, the directional carbon intensity $CI_h^{\uparrow}(\Delta)$ is the slope of that upward ramp direction chord; $CI_h^{\downarrow}(\Delta)$ is the slope of the chord in the downward ramp direction chord. The two chords agree with the tangent only while the ramp stays within the validity range of the LME, and they generally disagree with each other beyond it, because an upward ramp calls on the units above the margin in merit order while a downward ramp releases the units below it. Figure 1C shows the consequence in one hour of the real fleet (20 January 2025, 09:00 EPT, chosen as the hour whose error for a 100 MW ramp is closest to the median among hours with error above 10%). The marginal unit is coal unit 8102_1 and the LME is 0.96 t $CO_2$/MWh. A 100 MW upward ramp exhausts the headroom of that unit after 47.5 MW and is met for the remaining 52.5 MW by gas combined-cycle unit 56963_E101, the next unit in merit order, which is more expensive but cleaner at 0.40 t $CO_2$/MWh, so the emission the ramp causes is 0.66 t $CO_2$/MWh and the LME overstates it by 45%. A 100 MW downward ramp stays within unit 8102_1, so the emission it avoids is 0.96 t $CO_2$/MWh, equal to the LME. In this hour the LME is exact for ramping down and wrong by 45% for ramping up, and the two directions differ by a factor of 1.45. The direction of the error is not fixed: the LME understates the ramp when the next unit is dirtier than the marginal one and overstates it when the next unit is cleaner, as here.

### The marginal signal misstates large ramps in most hours, and most often in winter

Figure 2A shows the share of hours in which the LME misstates the emission of an upward ramp by more than 10% for every month of 2025, with every committed unit free to respond (full flexibility). For a 50 MW ramp the share is 7.14% to 11.56% of hours and for 100 MW it is 15.00% to 21.24%, with no seasonal pattern. From 300 MW upward the share grows with ramp size and varies with season: 28.49% to 50.00% of hours at 300 MW and 41.80% to 83.20% at 1 GW. In the other eleven months of 2025 the LME is wrong by more than 10% for a 1 GW ramp in 60.00% to 83.20% of hours; July, at 41.80%, is the exception. The magnitude of the error is also large. For a 1 GW ramp, the median relative error of the LME over all hours is 17.35% to 33.42% in every month except July, where it is 5.83%; conditional on exceeding 10%, the median error is 37.19% in January and 29.98% in July, and the 90th percentile over all hours is 58.10% and 44.41%.

The seasonal pattern follows from which units meet the ramp (Figure 2B). In winter season (January, February, March and December), a 1 GW ramp is met almost entirely (94.56% to 97.43%) by coal (44.10% to 47.27%) and gas combined-cycle units (47.68% to 53.33%), whose emission rates (0.90 and 0.38 t $CO_2$/MWh at the fleet medians) are far apart, so every breakpoint the ramp crosses changes the marginal rate by a large step. In July, gas combustion turbines and gas steam units, at 0.57 and 0.61 t $CO_2$/MWh at the fleet medians, supply 30.28% of the ramp, compared with 2.57% to 5.44% in those four months and at most 16.96% in any other month. In July a 1 GW ramp is met by more units (a median of 9 units per hour against 6 in January) whose emission rates lie closer together. As the ramp moves up the merit order, the marginal rate jumps each time the next unit takes over, and the size of these jumps measures how uneven the staircase of Figure 1C is. In the median July hour the mean jump between consecutive responding units is 0.127 t $CO_2$/MWh against 0.249 in January, the capacity-weighted spread of the responding units' emission rates is 0.127 against 0.230 t $CO_2$/MWh, and a jump larger than 0.3 t $CO_2$/MWh occurs somewhere inside the ramp in 66.94% of July hours against 90.59% of January hours (Figure S1). The signal does not work better in summer; the fleet's emission rates happen to lie closer together.

Limiting each unit's response to the largest change it made in any hour of the month (the ramp-limited case) raises the share of hours with error above 10% in every case: from 20.30% to 24.06% at 100 MW and from 83.20% to 87.50% at 1 GW in January, and from 16.94% to 18.01% and from 41.80% to 46.91% in July. Tighter limits raise it further (Table S1), so the full-flexibility case is a lower bound on how often the LME fails for a given fleet. The downward ramp shows a similar pattern (Table S2). Within the day, the LME fails more often when load is low: at 1 GW over the year, in 70.64% of hours between midnight and 06:00 against 59.01% between 17:00 and 21:00, and in 73.12% of weekend hours against 65.53% of weekday hours (Figure S2), because at low load the committed fleet is nearer the boundary between gas and coal units and the marginal unit has less headroom.

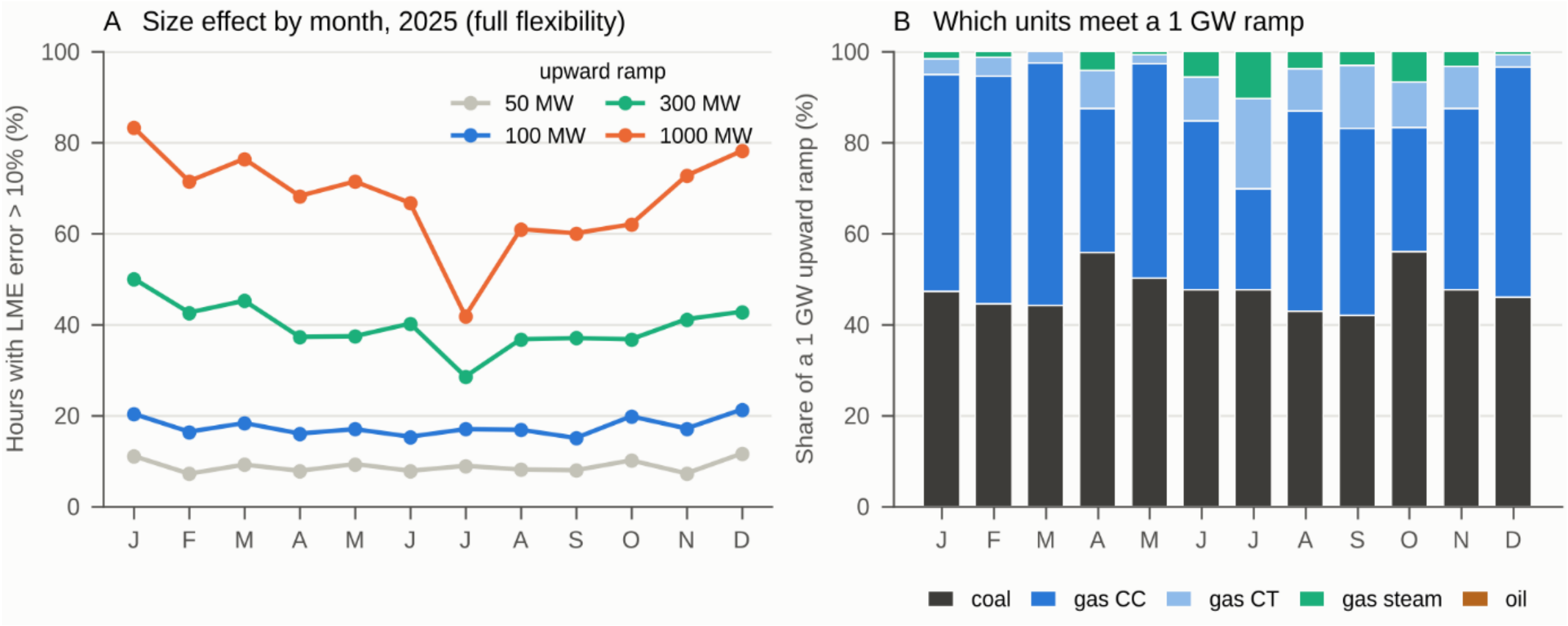

Figure 2. The LME misstates large ramps in most hours of 2025, and most often in winter. (A) Share of hours in which the LME misstates the emission of an upward ramp by more than 10%, by month and ramp size, full flexibility. (B) Share of a 1 GW upward ramp met by each fuel class, by month.

## Upward and downward ramps incurred emissions differ hour by hour but agree on average

Figure 3A plots, for every hour of January and July, the emission a 100 MW upward ramp causes against the emission a 100 MW downward ramp avoids. Each point is one hour. A point lies on the diagonal when the same unit absorbs the whole ramp in both directions, so that both intensities equal that unit's emission rate. A point lies off the diagonal when the upward and downward ramps are met by different units: an upward ramp is served by the marginal unit and then by the next units above it in merit order, while a downward ramp is absorbed by the marginal unit and then by the units below it. In the hour of Figure 1C, for example, the downward ramp stays within the marginal coal unit (0.96 tCO2/MWh) while the upward ramp runs into a gas combined-cycle unit (0.40 tCO2/MWh), so the point lies below the diagonal at (0.96, 0.66). The points form a lattice because the measured emission rates of the fleet fall into three groups set by fuel carbon content and heat rate: the median rate is 0.38 t $CO_2$/MWh for gas combined-cycle units (10th to 90th percentile 0.35 to 0.59), 0.57 for gas combustion turbines (0.46 to 0.83), 0.61 for gas steam units (0.51 to 0.91) and 0.90 for coal units (0.82 to 1.03). Each hourly intensity is the rate of one responding unit or a capacity-weighted average of multiple responding units, so the values concentrate at these levels and between them. In full flexibility case, the two intensities differ by more than 10% of their mean in 37.63% of hours in January and 25.67% in July at 100 MW, rising to 65.19% and 42.88% at 1 GW (Figure 3B); over the twelve months the share at 100 MW ranges from 25.67% to 37.63% and at 1 GW from 42.88% to 76.75%; they differ by more than 25% in 28.49% and 15.99% of hours at 100 MW. In the ramp-limited case, two intensities differ by more than 10% at 100 MW are 42.34% and 26.08%.

Over the month the difference disappears. The monthly means of $CI^{\uparrow}$ and $CI^{\downarrow}$ differ by at most 0.010 t $CO_2$/MWh in January and July (Figure 3C) and 0.025 t $CO_2$/MWh in any month of 2025 at full flexibility, and 0.014 t $CO_2$/MWh in the ramp-limited case at every ramp size (Figure 3C). Direction therefore matters for a decision taken in one hour but is invisible in a monthly or annual average, which explains why a regression of monthly emissions on load, the usual way of estimating marginal emission factors from data cannot detect it.

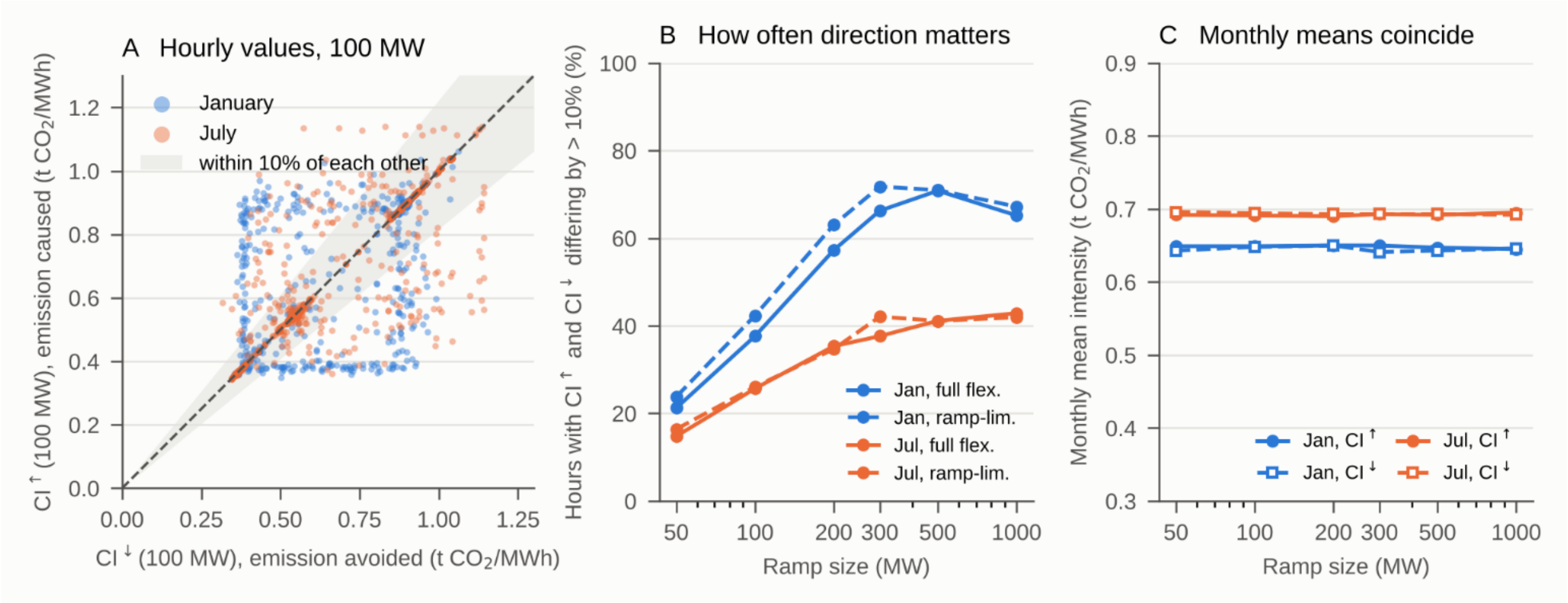

Figure 3. Upward and downward ramps differ hour by hour but agree on monthly average. (A) Hourly upward against downward intensity, 100 MW. (B) Share of hours in which the two differ by more than 10%, by ramp size. (C) Monthly means coincide.

## Location-based reporting assigns other consumers close to half of a ramp's emission

Figure 4A compares the regional average intensity $a_h$ with the emission a 100 MW upward ramp causes hour by hour. In 99.60% or more of hours the ramp causes more per MWh than the average assigns, because the average includes the nuclear, hydro and renewable generation that does not respond to a ramp but dilutes the average emission rate, while the responding units are gas and coal. The mean of $a_h$ is 0.35 t $CO_2$/MWh in January and 0.36 t $CO_2$/MWh in July; the mean of $CI^{\uparrow}$(100 MW) is 0.65 t $CO_2$/MWh and 0.69 t $CO_2$/MWh, respectively.

Summed over the month, a 100 MW ramp held in every hour causes 48.25 kt CO2 in January and 51.42 kt in July, of which location-based reporting assigns the campus 25.78 kt and 26.93 kt (Figure 4B). The difference, 22.47 kt in January and 24.49 kt in July, is emission that the campus's ramps cause but other consumers declare. Expressed as a share of the caused emission, the accounting gap $g$(100 MW) of Experimental Procedures is 46.58% in January and 47.63% in July at full flexibility, and 42.52% and 44.37% in the ramp-limited case. Over the twelve months of 2025, $g$(100 MW) at full flexibility ranges from 46.58% in January to 58.64% in April (Figure 4C): the gap is largest in spring and autumn, when low load and high nuclear and renewable shares pull the regional average down to 0.27 to 0.30 t $CO_2$/MWh while the responding units remain gas and coal. January and July, the two months with the highest regional averages, are the two months with the smallest gap. $g(\Delta)$ takes the same value at every ramp size from 50 MW to 1 GW (in every month the values at 50 MW and at 1 GW differ by at most 1.42% of the caused emission, Figure 4C), because both the caused and the declared emission scale with the ramp and the ratio depends only on how far the responding units' emission rates sit above the regional average. A downward ramp is under-credited by the same share (Table S2). The emission a campus's ramps cause but do not declare is not lost. It raises the regional average, so every other consumer in the region declares a share of it in proportion to its consumption.

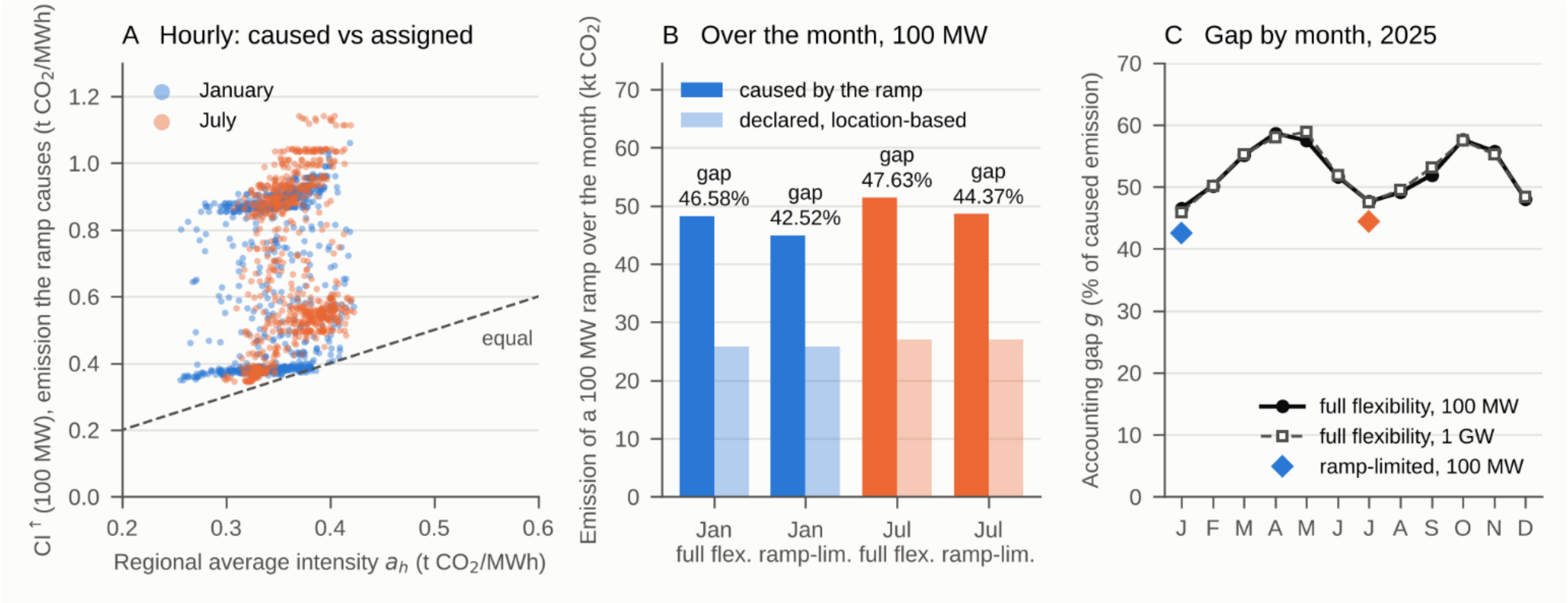

Figure 4. Location-based reporting assigns other consumers about half of the emission a ramp causes. (A) Hourly regional average intensity against the emission a 100 MW upward ramp causes, January and July, full flexibility. (B) Over the month: emission caused by a 100 MW ramp held in every hour against the emission declared under location-based reporting. (C) Accounting gap g by month for 2025 (full flexibility, 100 MW and 1 GW), with the ramp-limited January and July values as markers.

## PJM's published signal persists only weakly between intervals

PJM publishes a five-minute marginal $CO_2$ rate for every node. Figure 5A classifies each interval at the Ashburn node, the Dominion zone and the PJM region by the state of the rate: negative, gas (0 to 0.45 t $CO_2$/MWh), intermediate (0.45 to 0.80 t $CO_2$/MWh) or coal (above 0.80 t $CO_2$/MWh). At Ashburn in January, 58.63% of intervals are in the gas state, 15.98% intermediate, 20.69% coal and 4.69% negative; in July, 43.39%, 23.86%, 27.89% and 4.86%. The shares at the three nodes differ by at most 3.04% of intervals in any state in either month, so Ashburn is representative of its zone and of the region. The negative and intermediate intervals, 20.67% in January and 28.72% in July, arise from congestion and from redispatch shared between units, and the merit-order model, which has no network, cannot produce them: it places 45.30% to 48.66% of hours in the coal state against PJM's 20.69% to 27.89%, none in the negative state, and in January only 4.70% in the intermediate state against PJM's 15.98% (Figure 5A, C). The hourly distribution of the model's LME is therefore not a substitute for the published signal, and the comparison in Figure 5C is a check on the model's fuel mix rather than a validation of individual hours.

The published signal changes state often. Consecutive five-minute intervals at Ashburn are in the same state 48.52% of the time in January and 46.00% in July; if the states were independent draws with the observed shares, the expected figures would be 41.44% and 32.53% (Figure 5B). The rank correlation between an interval's rate and the next is 0.13 in January and 0.18 in July. The signal therefore carries some persistence, more in July than in January, but a campus that reads the rate at the start of a five-minute interval has a less than even chance that the same fuel will be marginal in the next. For scheduling, this matters as much as the size effect: the LME is defined for an infinitesimal change and for the instant at which it is computed, and a ramp of hundreds of megawatts that lasts an hour violates both conditions.

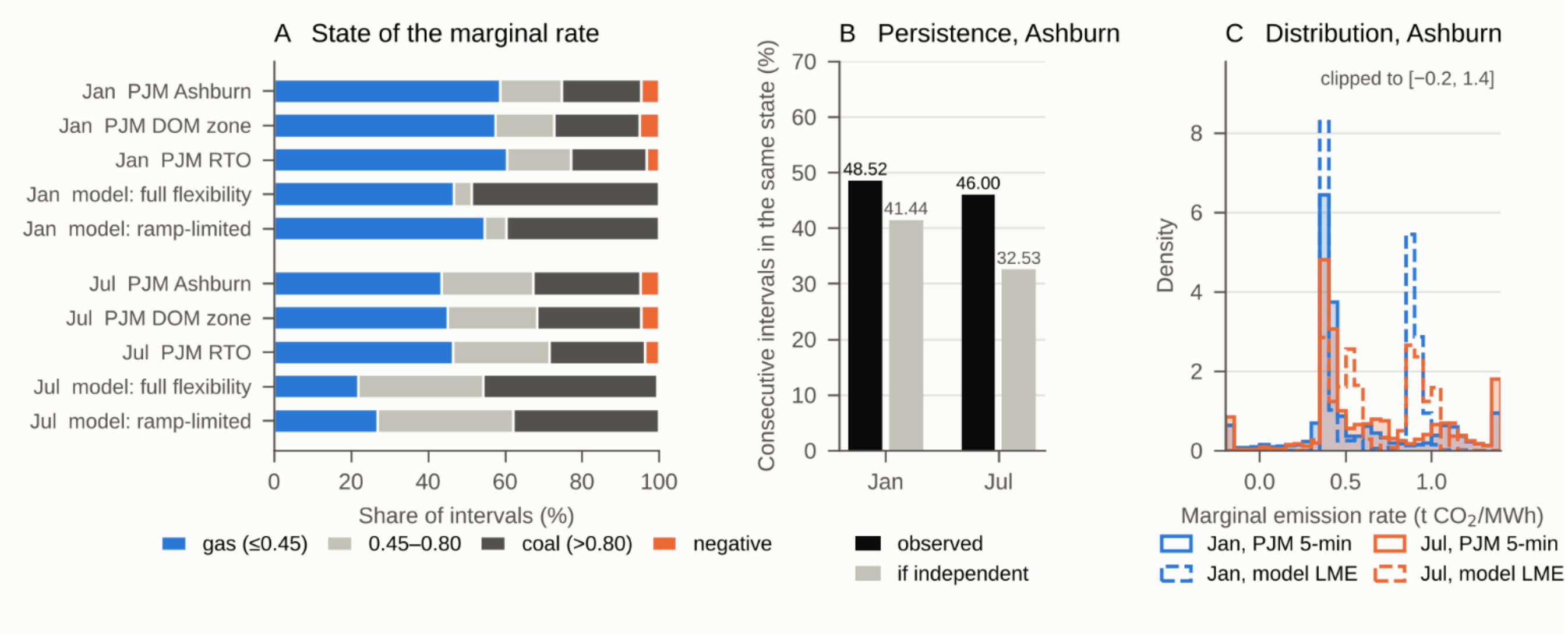

Figure 5. PJM's published five-minute marginal emission rate persists only weakly between intervals. (A) State of the marginal rate (negative, gas, intermediate, coal) at the Ashburn node, the Dominion zone and the PJM region, and in the model, January and July 2025. (B) Share of consecutive five-minute intervals in the same state at Ashburn against the share expected if states were independent. (C) Distribution of the rate at Ashburn against the model's hourly LME. Values clipped to [−0.2, 1.4] t CO2/MWh. All PJM quantities are derived statistics (PJM case 00342436).

## DISCUSSION

A carbon signal with a single value per location and interval cannot represent the emission a data-center ramp causes. On the real PJM fleet, the locational marginal emission rate misstates the emission of a 1 GW ramp by more than 10% in 60% to 83% of hours in eleven months of 2025, upward and downward ramps of the same size differ by more than 10% in 25.67% to 37.63% of hours at 100 MW and in 42.88% to 76.75% at 1 GW, and location-based reporting assigns 46.58% to 58.64% of the emission a ramp causes to other consumers. These results do not depend on how much flexibility the fleet is assumed to have, and they hold in every month of 2025. They have a single cause. A finite ramp is not served by one unit: as it grows, it exhausts the headroom of the marginal unit and passes to the next unit in merit order, and then to the next. Each time it passes to a new unit, the marginal emission rate changes to that unit's rate. On the PJM fleet these rates fall into three groups, with medians of 0.38, 0.57 to 0.61 and 0.90 t CO2/MWh (Table S3), so a ramp that passes from a gas combined-cycle unit to a coal unit changes the marginal rate by more than a factor of two.

### What a campus should schedule on, and what it should report

For scheduling, the right carbon signal depends on the action the campus is deciding. A DC campus deciding whether to add 50 MW in the next hour can use the LME: it is within 10% of the caused emission in 88% to 93% of hours. A campus deciding whether to add or shed 300 MW or more cannot rely on the LME in most hours, and the error is not a bias that a correction factor would remove. It changes sign and size from hour to hour with the headroom of the marginal unit (Figure 1C), and it differs between ramping up and ramping down in the same hour (Figure 3A). The quantity the campus needs is the directional carbon intensity for the size and direction of the ramp it is considering, which a grid operator can compute from the same dispatch model it already uses to publish the LME: the LME is the slope at the operating point, and the directional intensity is the slope of the chord to the end of the ramp. Publishing the intensity for a few standard ramp sizes in each direction, such as 100, 200,…,900 and 1,000 MW in ladder style, would cost the operator one additional dispatch solution per size and direction per interval.

For campus emission reporting, the finding bears on a live question. The Greenhouse Gas Protocol is revising the Scope 2 standard, with a consolidated public consultation planned for the second quarter of 2027 and publication in 2028 [21]. The debate so far has concerned the market-based method: whether contractual instruments must be matched hourly and delivered to the consumer's grid. Our results concern the location-based method, which the revision keeps: it assigns a campus the regional average intensity, and on PJM in 2025 the monthly regional average is 0.27 to 0.36 t $CO_2$/MWh while the emission a ramp causes averages 0.65 to 0.69 in January and July. The location-based figure is not wrong as an allocation; it allocates all generation emissions to all consumption once. It is the wrong answer to the question a campus operator, a regulator or a neighbor asks, which is how much emission the campus's decisions caused. The two numbers should be reported side by side, as Gorka et al. argued from a test system [11] and as we find on the real fleet, because a campus that schedules on the consequential signal and reports on the attributional one has an incentive to reduce its reported emissions by shifting them onto other consumers rather than by reducing them.

### Relation to earlier work

The validity ranges of Nilges et al. show when the LME stops being exact [12]; the directional intensity tells it what the emission is once the LME has stopped being exact, and the shares in Figure 2 tell it how often that happens. The two are complementary and the ramp-limited case of our model gives validity ranges of the same order as theirs: mostly tens of megawatts, and in some hours a few megawatts. Cote and Sun [14] found that the LME predicted the emission of 1,000 random 200 MW siting decisions to within 14% in sum; Figure 2A shows that at 200 MW the LME is wrong by more than 10% in 24% to 37% of individual hours, and Figure 3C shows the monthly means of the upward and downward intensities within 0.010 t $CO_2$/MWh of each other in January and July while the hourly values differ in 25.67% to 37.63% of hours at 100 MW. Both studies are right about the quantity they report. A sum or a mean is the right quantity for a siting or procurement decision that commits a load for planning decision; an hourly value is the proper quantity for a scheduling decision, and the two quantities disagree. Khanpour et al. [13] bound the exceedance of the LME under uncertain demand by the probability that the binding constraints change; a deliberate ramp with known size and direction has no such uncertainty, and the directional intensity replaces the bound with the exact value.

### Limitations

The dispatch problem employs a copper plate model. The fleet is dispatched in merit order against total fossil demand, so every committed unit can respond to a ramp at Ashburn regardless of where it stands. A network removes units from the responding set when lines are congested, introduces breakpoints at line limits as well as at unit limits, and produces the negative and intermediate marginal rates that make up 20.67% of PJM's published intervals in January and 28.72% in July (Figure 5A) and that the model cannot reproduce. Each of these effects adds breakpoints to the emission curve without removing any, which is why Nilges et al. found validity ranges of about 5 MW in a congested grid and why we expect, but have not shown, that the shares in Figure 2 are lower bounds for the networked system. The theory for the networked case, including negative intensities and sign reversal, is given in [15] on 30-bus and 118-bus systems; applying it to PJM requires a network model that is not public. The same limitation means that the comparison with PJM's published signal in Figure 5 tests the model's fuel mix and not its hour-by-hour values: the model places 45.30% to 48.66% of hours in the coal state against PJM's 20.69% to 27.89%, and a reader should take the model's absolute shares as those of a merit-order fleet, not of PJM's dispatch.

Three further simplifications are stated in Experimental Procedures and affect the numbers rather than the conclusions. Each unit has one emission rate for the month, so part-load efficiency is averaged out; the daily coal-gas spread is calibrated to observed coal output, so the fuel mix in Figure 2B is partly a product of the calibration; and imports, exports and the units in North Carolina are omitted. The study covers one year of one interconnection. The mechanism is general: a ramp crosses between groups of units with different emission rates, and its size in another system depends on how far apart that system's emission rates lie: Figure 2 shows the effect halving in July when gas turbines and steam units, with rates between those of combined cycle and coal, supply a larger share of the ramp. A fleet of a single fuel would show a small size effect; a more diverse fleet of hydro, gas and coal, as in several Asian and South American systems, would show a larger one.

### Implications for decarbonization

The size effect is a property of the fleet's emission rates, not of the data center. It will shrink as coal retires and the rates of the remaining units converge, and it will grow wherever a low-carbon unit sits next to a high-carbon one in merit order. This reveals a question under-investigated: the optimal siting and sizing of clean generation that could respond to directional large load variation in the merit order. A coal phase-out schedule that leaves coal as the marginal unit at the nodes where data centers connect would reduce average emissions without reducing the emissions that data-center ramps cause, and location-based reporting would not show the difference. A second open question is on-site generation. Where campus capacity outgrows what the grid can connect, operators are adding behind-the-meter generation, and a campus that both consumes and generates changes the sign and the size of its ramps as seen by the grid; how its emission should be accounted for and reported when it is a producer as well as a consumer is not settled by the present framework [6].

## EXPERIMENTAL PROCEDURES

### Resource availability

*Lead contact.* Further information and requests for resources should be directed to and will be fulfilled by the lead contact, Xiaoying Shi (x1shi@polyu.edu.hk).

*Materials availability.* This study did not generate new materials.

*Data and code availability*. Hourly unit-level generation, heat input and $CO_2$ emissions are from the U.S. Environmental Protection Agency Clean Air Markets Program Data (CAMPD) [16]. Plant locations and balancing-authority assignments are from the U.S. Energy Information Administration Form EIA-860 (2025) [17]. Hourly generation by fuel for all of 2025 and the five-minute nodal marginal emission rates for January and July 2025 are from PJM Data Miner 2 [18] and are used with PJM's permission for the publication of derived results (PJM case 00342436); they are not redistributed with this article and can be obtained from PJM Data Miner 2 with the download scripts in the code archive. The code that reproduces all results and figures is available at Zenodo, https://doi.org/10.5281/zenodo.24074188 (version 1.0).

## Directional carbon intensity

Let $E_h(\theta)$ be the $CO_2$ emission of the system dispatch in hour $h$ when the campus draws θ (MW):

$$E_h(\theta) = \sum_k e_k\, p^*_{k,h}(\theta) \tag{1}$$

where $e_k$ is the emission rate of unit $k$ ($tCO_2$/MWh), $p^*_{k,h}(\theta)$ is the output of unit $k$ (MW) in the least-cost dispatch of hour $h$ that meets the observed fossil demand plus θ (the dispatch problem is given below), and the sum runs over all fossil units. $E_h$ is in $tCO_2$/MWh per hour. It is piecewise linear in θ: within each piece the same unit is marginal, and at each breakpoint the marginal unit changes.

The directional carbon intensities of a campus ramp of size Δ in hour $h$ are

$$CI_h^{\uparrow}(\Delta) = \frac{E_h(\theta_0+\Delta) - E_h(\theta_0)}{\Delta}, \quad CI_h^{\downarrow}(\Delta) = \frac{E_h(\theta_0) - E_h(\theta_0-\Delta)}{\Delta} \tag{2}$$

where $E_h(\theta)$ is the $CO_2$ emission of the system dispatch in hour $h$ ($t/h$) when the campus draws θ (MW), $\theta_0$ is the campus baseline load (MW), and $\Delta>0$ is the ramp power (MW). $CI_h^{\uparrow}(\Delta)$ is the emission caused per MWh of an upward ramp and $CI_h^{\downarrow}(\Delta)$ is the emission reduced per MWh of a downward ramp, respectively; both in $tCO_2$/MWh. The one-sided locational marginal emissions are

$$\lambda_h^{\uparrow} = \frac{E_h(\theta_0+\varepsilon) - E_h(\theta_0)}{\varepsilon}, \quad \lambda_h^{\downarrow} = \frac{E_h(\theta_0) - E_h(\theta_0-\varepsilon)}{\varepsilon} \tag{3}$$

where ε=0.1 MW. As $E_h$ is piecewise linear in θ, $CI_h^{\uparrow}(\Delta) = \lambda_h^{\uparrow}$ for every Δ within the upward validity range and $CI_h^{\downarrow}(\Delta) = \lambda_h^{\downarrow}$ within the downward validity range, and each directional intensity equals the one-sided LME plus one correction term per breakpoint of $E_h$ crossed by the ramp [15].

## Fleet and data

The model is run month by month for the twelve months of 2025, and the fleet is rebuilt for each month. The fossil fleet of a month comprises the CAMPD units in the PJM balancing authority (EIA-860) with positive output in that month, located in Delaware, Illinois, Indiana, Kentucky, Maryland, Michigan, New Jersey, Ohio, Pennsylvania, Virginia and West Virginia: 682 to 720 units depending on the month (682 in January: 89 coal units, 39.6 GW; 191 gas combined-cycle units, 61.2 GW; 298 gas combustion turbines, 27.1 GW; 40 gas steam units, 6.8 GW; 58 oil units, 3.3 GW; 6 other fossil units, 0.3 GW; the additional units in summer are gas steam and oil peaking units). CAMPD units in North Carolina were not included. For each unit $k$ and month, the emission rate $e_k$ ($tCO_2$/MWh) and heat rate $r_k$ (MMBtu/MWh) are the month's totals of $CO_2$ mass and heat input divided by gross generation; units with 100 MWh or less in the month take the median of their class. The maximum output $\overline{p}_k$ is the largest hourly gross output observed in the month, and the minimum output $\underline{p}_k$ is the 5th percentile of its hourly output in the hours it ran, capped in each hour at the output observed in that hour. A unit is committed in hour $h$ if its observed gross output in that hour is positive. Non-fossil generation (nuclear, hydro, wind, solar, other renewables and storage) is taken from PJM hourly generation by fuel for each month and held at its observed value.

## Dispatch model

In the copper plate model, committed fossil units meet the observed fossil output plus the campus load at least cost in each hour $h$:

$$
\begin{aligned}
&\min_p \sum_k c_{k,d}\, p_k \\
&\text{s.t.} \quad \sum_k p_k = D_h + \theta, \\
&\qquad \theta = \theta_0 + \Delta \\
&\qquad \underline{b}_{k,h} \le p_k \le \overline{b}_{k,h},
\end{aligned} \tag{4}
$$

where $p_k$ is the output of unit k (MW), $c_{k,d}$ is its marginal cost on day $d$ (\$/MWh), $D_h$ is the observed total gross output of the fossil fleet in hour $h$ (MW), θ is the DC campus load (MW), with $\theta_0$ is the DC campus capacity, Δ is its ramping power. $\underline{b}_{k,h}$ and $\overline{b}_{k,h}$ are the output bounds of unit $k$ in hour $h$ (MW), both zero for uncommitted units. This is a linear programming problem is solved exactly by loading committed units in order of increasing cost above their minimum outputs. The emission of the dispatch is $E_h(\theta) = \sum_k e_k p_k^*$, where $p_k^*$ is the optimal output. The marginal cost is

$$c_{k,d} = r_k \pi_{f(k),d} + v_{f(k)}, \tag{5}$$

where $\pi_{f(k),d}$ is the price of the fuel $f(k)$ of unit $k$ on day $d$ in \$/MMBtu and $v_{f(k)}$ is the variable operating and maintenance cost of its class in \$/MWh. We use PJM's default VOM adders for cost-based offers, the sum of the minor-maintenance and operating-cost defaults effective 1 January 2026 [19]: 1.77 for combined cycle, 5.57 for simple-cycle combustion turbines and 5.89 for fossil steam units, applied to coal, gas steam, oil and other fossil units. Coal and oil prices are fixed at 2.48 and 16.63 \$/MMBtu, the 2025 average delivered cost of coal and petroleum liquids to electric utilities [20]. The gas price of each day is a calibrated parameter, found by bisection on the search interval [0.5, 25] \$/MMBtu (30 iterations), so that modelled coal output over the 24 hours of the day equals observed coal output. The calibrated price lies between 2.63 and 4.06 \$/MMBtu over the two months (2.91 to 4.06 in January, 2.63 to 4.01 in July), the same range as the daily Henry Hub spot price, which EIA reports between 2.65 and 9.86 \$/MMBtu in 2025 [22]; the search interval is set wide enough to contain both and to allow for the higher delivered prices at PJM trading points on cold days. Because the coal-gas order of the merit list is set by this calibration, the results depend only weakly on the fixed cost parameters: halving or raising all variable costs by 50%, setting them to a flat 3 \$/MWh, moving the coal price to 2.0 or 3.0 \$/MMBtu, or the oil price to 10 or 25 \$/MMBtu changes the share of January hours with LME error above 10% by at most 1.6% of hours at 100 MW and 1.3% at 1 GW, and the accounting gap g(100 MW) by at most 1.3% of the caused emission (Table S4). Identical costs are separated by a tie-breaking term below $10^{-4}$ \$/MWh.
The campus base load $\theta_0$ is assumed to be 300 MW for ramps Δ of 50, 100 and 200 MW and 1,100 MW for ramps of 300, 500 and 1,000 MW, respectively; so that every downward ramp stays within the base. The observed fossil output already serves all existing load, including the data centers operating in 2025. The campus is added on top of it as one further load, with base load $\theta_0$ and ramps of ±Δ so that the results describe the effect of a new campus or a new building rather than of the existing fleet of data centers. The DC campus load takes the values $\theta_0$, $\theta_0+\Delta$, $\theta_0-\Delta$, which give $E_h(\theta_0)$, $E_h(\theta_0+\Delta)$ and $E_h(\theta_0-\Delta)$, respectively.

## Flexibility of unit response

Two cases bound the response of the fleet. In the full-flexibility case, the bounds are $\underline{b}_{k,h} = \underline{p}_k$ and $\overline{b}_{k,h} = \overline{p}_k$ for every committed unit, so any committed unit can move across its full range within the hour and the dispatch is set by merit order alone. In the ramp-limited case, the dispatch is anchored at the observed output and each committed unit may move from it by no more than the largest allowable change it made in one hour during the month:

$$
\begin{aligned}
\underline{b}_{k,h} &= \min\big(o_{k,h}, \max(\underline{p}_k, o_{k,h} - \beta_k)\big), \\
\overline{b}_{k,h} &= \min\big(\max(\overline{p}_k, o_{k,h}), o_{k,h} + \beta_k\big)
\end{aligned} \tag{6}
$$

where $o_{k,h}$ is the observed output of unit $k$ in hour $h$ (MW) and $\beta_k$ is the largest absolute change in its output between consecutive hours in which it ran in both (MW); $\beta_k = 0$ for units with five or fewer such hour pairs. Figure S3 shows the hour of Figure 1C under both assumptions. The full-flexibility case is run for all twelve months; the ramp-limited case and the sensitivity below are run for January and July. To test sensitivity, $\beta_k$ is replaced by the $q_{th}$ percentile of the same changes for $q$ from 50 to 99 (Table S1). For $q \le 70$ the bands are too tight to be physical: in some hours the committed fleet cannot meet a 1,100 MW campus plus a ramp of 300 MW or more within the bands, although the observed total fossil output of the same fleet changed by more than 2.1 GW between consecutive hours in 36.34% of January hours and 60.16% of July hours (median change 1.56 and 2.55 GW). Those bands are therefore rejected rather than reported, and the ramp-limited case uses the largest observed change ($q$=100), which is the tightest band consistent with what the fleet is observed to do. We also tested whether PJM's five-minute signal could select $q$, by minimising the Wasserstein-1 distance between the model's hourly upward LME and PJM's rate at Ashburn, both clipped to [0, 1.5] $tCO_2$/MWh. The distance fell with small fluctuations, from 0.148 $tCO_2$/MWh at $q$=100 to 0.106 $tCO_2$/MWh at $q$=50, with no minimum inside the range, and stayed above 0.1 $tCO_2$/MWh for every band (Figure S4).

The model produces no negative and few intermediate (0.45 to 0.80 $tCO_2$/MWh ) marginal rates, whereas these make up 20.67% of PJM's five-minute intervals at Ashburn. The signal therefore does not identify $q$ and was not used to choose it.

## Error, direction and accounting metrics

The relative error of the LME for a ramp of size Δ in hour $h$ is:

$$\eta_h^{\uparrow}(\Delta) = \frac{\left|\lambda_h^{\uparrow} - CI_h^{\uparrow}(\Delta)\right|}{\left|CI_h^{\uparrow}(\Delta)\right|} \tag{7}$$

and likewise for the downward direction. Hours with $|CI_h(\Delta)|<0.05 tCO_2/MWh$ would be excluded, because a relative error is not meaningful when the ramp causes almost no emission (an absolute difference of 0.01 $tCO_2$/MWh would register as a 100% error against a 0.01 denominator). In practice no hour falls below the threshold: the smallest $|CI_h^{\uparrow}(\Delta)|$ in 2025 is 0.24 $tCO_2$/MWh, so the shares reported are over all hours. We report the share of hours with $\eta_h(\Delta)>0.1$. The directional difference relative to their mean in hour $h$ is

$$\delta_h(\Delta) = \frac{\left|CI_h^{\uparrow}(\Delta) - CI_h^{\downarrow}(\Delta)\right|}{\frac{1}{2}\left(\left|CI_h^{\uparrow}(\Delta)\right| + \left|CI_h^{\downarrow}(\Delta)\right|\right)} \tag{8}$$

computed for hours in which the denominator is at least 0.05 $tCO_2$/MWh. The regional average intensity in hour $h$ is:

$$a_h = \frac{E_h(\theta_0)}{\sum_k p_k^* + G_h} \tag{9}$$

where $G_h$ is the non-fossil generation in hour h (MW), $E_h(\theta_0)$ is the emission of the fossil fleet at the operating point (t/h) and $\sum_k p_k^*$ is its output (MW). This is the location-based emission factor of the GHG Protocol Scope 2 Guidance applied at hourly resolution to the PJM balancing authority [4].

An upward ramp of size Δ in hour $h$ would cause the grid more emission $CI_h^{\uparrow}(\Delta)\Delta$ (t/h) than it without the ramp. Under location-based reporting the campus declares $a_h \cdot \Delta$ for the same electricity, because the method multiplies consumption by the regional average. The difference is $\left(CI_h^{\uparrow}(\Delta) - a_h\right)\Delta$ is emitted but not declared by the campus; it enters the regional average and is therefore declared, in proportion to their consumption, by all other consumers in the region. It is positive when the units that respond to the ramp emit more per MWh than the fleet average (coal responding in a fleet whose average includes nuclear and gas) and negative when they emit less.

To express the gap as a share, we assume that the campus ramps up by Δ in every hour of the month, so that the caused emission over the month is $\Delta\sum_h CI_h^{\uparrow}(\Delta)$ and the declared emission is $\Delta\sum_h a_h$. The share of the caused emission that location-based reporting does not assign to the campus is

$$g(\Delta) = \frac{\Delta\sum_h CI_h^{\uparrow}(\Delta) - \Delta\sum_h a_h}{\Delta\sum_h CI_h^{\uparrow}(\Delta)} = 1 - \frac{\sum_h a_h}{\sum_h CI_h^{\uparrow}(\Delta)} \tag{10}$$

where both sums run over the hours of the month with $CI_h^{\uparrow}(\Delta) > 0.05$ $tCO_2$/MWh; $g$ depends on Δ only through $CI_h^{\uparrow}(\Delta)$. A value of $g$=0.4 means that the campus declares 60% of the emission its ramps caused and the remaining 40% is declared by other consumers. Hours in which the ramp causes almost no emission are excluded because the ratio is uninformative there. Summing over the month rather than taking the median of the hourly ratios matches how Scope 2 emissions are reported, as a total over a reporting period, and lets hours with negative differences offset hours with positive ones. $g$ is a property of the two intensities over the month, not of a particular campus schedule: a campus that ramps only in some hours would obtain the same ratio evaluated over those hours. The downward case, in which the campus is credited $a_h \cdot \Delta$ for avoiding $CI_h^{\downarrow}(\Delta)\Delta$, is defined in the same way and is reported in the Supplemental Information.

## PJM published marginal emission rate

PJM's five-minute nodal marginal $CO_2$ rate for January and July 2025 was obtained for the Ashburn node (pnode 34886139), the Dominion zone (34964545) and the PJM regional transmission organization (1) and converted from lb/MWh to t/MWh by the factor 0.000453592. All three nodes lack the same intervals: 16 of 288 on 22 January, 40 of 288 on 23 January, 6 to 8 of 288 on 9 July and 1 to 2 on five other July days. Each interval is assigned a state: negative (≤0), gas (0 to 0.45 t/MWh), intermediate (0.45 to 0.80) or coal (>0.80). Persistence is the share of consecutive intervals with the same state, and the share expected by chance is $\sum_s P_s^2$ , where $P_s$ is the share of intervals in state $s$. The same state thresholds are applied to the model's hourly $\lambda_h^{\uparrow}$.

## Software

Data processing and analysis used Python 3.12.10 with pandas 3.0.6, NumPy 2.5.3, SciPy 1.18.1 and Matplotlib 3.11.2; the code archive lists the package versions.

## Acknowledgments

The authors would like to acknowledge the Hong Kong Polytechnic University for the financial support (P0052766-Postdoc Matching Fund Scheme 23/24 4th round, P0047700 - International Centre of Urban Energy Nexus (UEX), P0051003 - Using Big Mobility Data for Smart Charging/Discharging Planning and Adaptation for Hong Kong's Smart Mobility System through Transfer Learning), the State Key Laboratory of Climate Resilience for Coastal Cities at the Hong Kong Polytechnic University and the Strategic Topics Grant (STG2/P-705/24-R) of the Hong Kong Research Grants Council.

## Author contributions

Conceptualization, X.S.; Methodology, X.S.; Software, X.S.; Formal analysis, X.S.; Data curation, X.S.; Writing – original draft, X.S.; Writing – review & editing, X.S. and J.Y.; Supervision, J.Y.; Funding acquisition, J.Y.

## Declaration of interests

The authors declare no competing interests.

## Declaration of generative AI and AI-assisted technologies in the writing process

During the preparation of this work the authors used Claude (Anthropic) to draft and edit the text of the manuscript and the supplemental information and to assist in writing and documenting the analysis and plotting code in the public code archive. After using this tool, the authors reviewed and edited the content, independently re-ran all computations, and take full responsibility for the content of the publication.

## Supplemental information

Document S1. Figures S1 to S4, Tables S1 to S4 and Notes S1 and S2.

# Supplemental Information

One number per hour is not enough: directional carbon intensity of data-center ramps on the PJM fleet

Xiaoying Shi and Jinyue Yan

## Supplemental figures

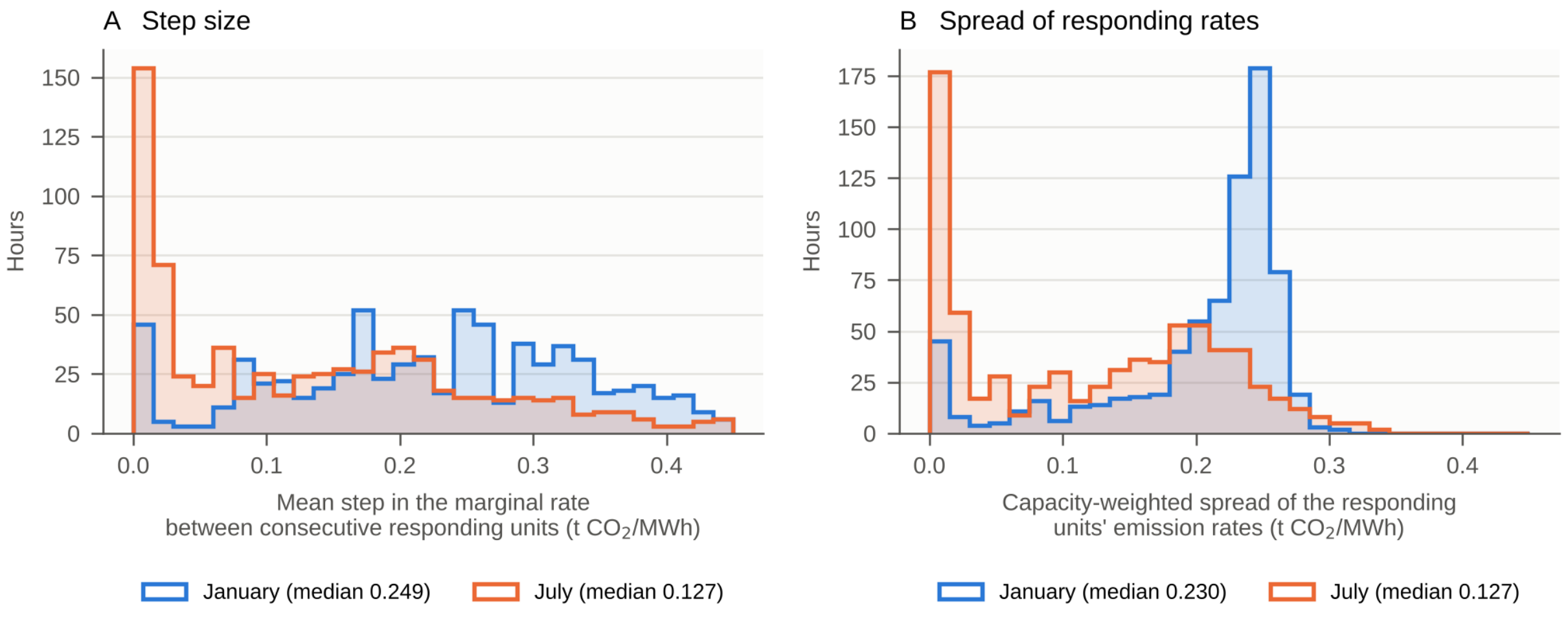

**Figure S1. A 1 GW ramp climbs a steeper staircase in January than in July, related to Figure 2.** For every hour of January and July 2025 (full flexibility), the units that respond to a 1 GW upward ramp are listed in merit order and the change in the marginal emission rate between consecutive units is recorded (one such staircase is shown in Figure 1C). (A) Distribution over hours of the mean step within the hour. (B) Distribution over hours of the capacity-weighted standard deviation of the responding units' emission rates. Medians are given in the legends. A 1 GW ramp is met by a median of 6 units per hour in January and 9 in July, and a step larger than 0.3 t $CO_2$/MWh occurs inside the ramp in 90.59% of January hours against 66.94% of July hours. Source: results/ramp_steps.csv (per hour) and results/ramp_steps_summary.csv (real/ramp_steps.py).

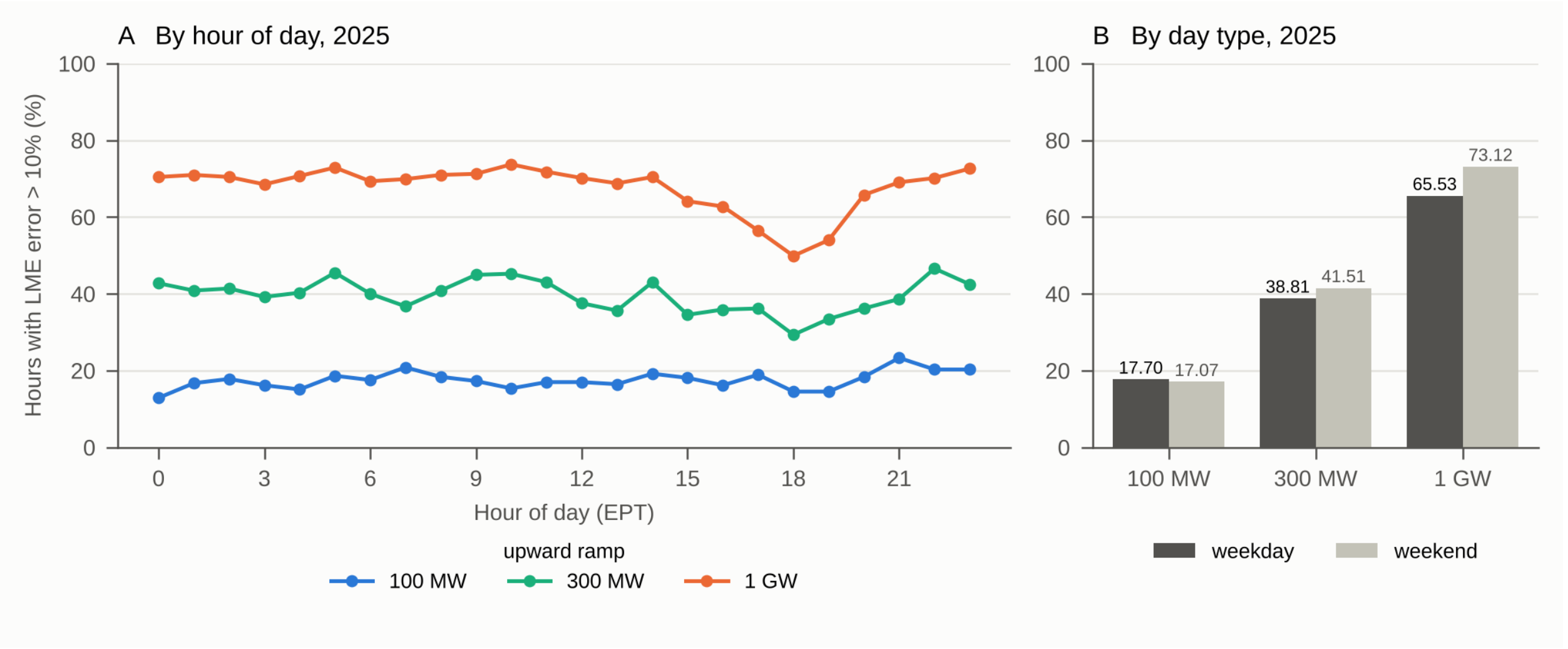

**Figure S2. The LME fails more often at low load, related to Figure 2.** Share of hours of 2025 in which the LME misstates an upward ramp by more than 10%, full flexibility, (A) by hour of day and (B) by weekday

or weekend, for ramps of 100 MW, 300 MW and 1 GW. For a 1 GW ramp the share is 70.64% between 00:00 and 05:59 against 59.01% between 17:00 and 21:59, and 73.12% on weekends against 65.53% on weekdays. Source: results/figS2_daytype.csv (real/figS2_daytype.py).

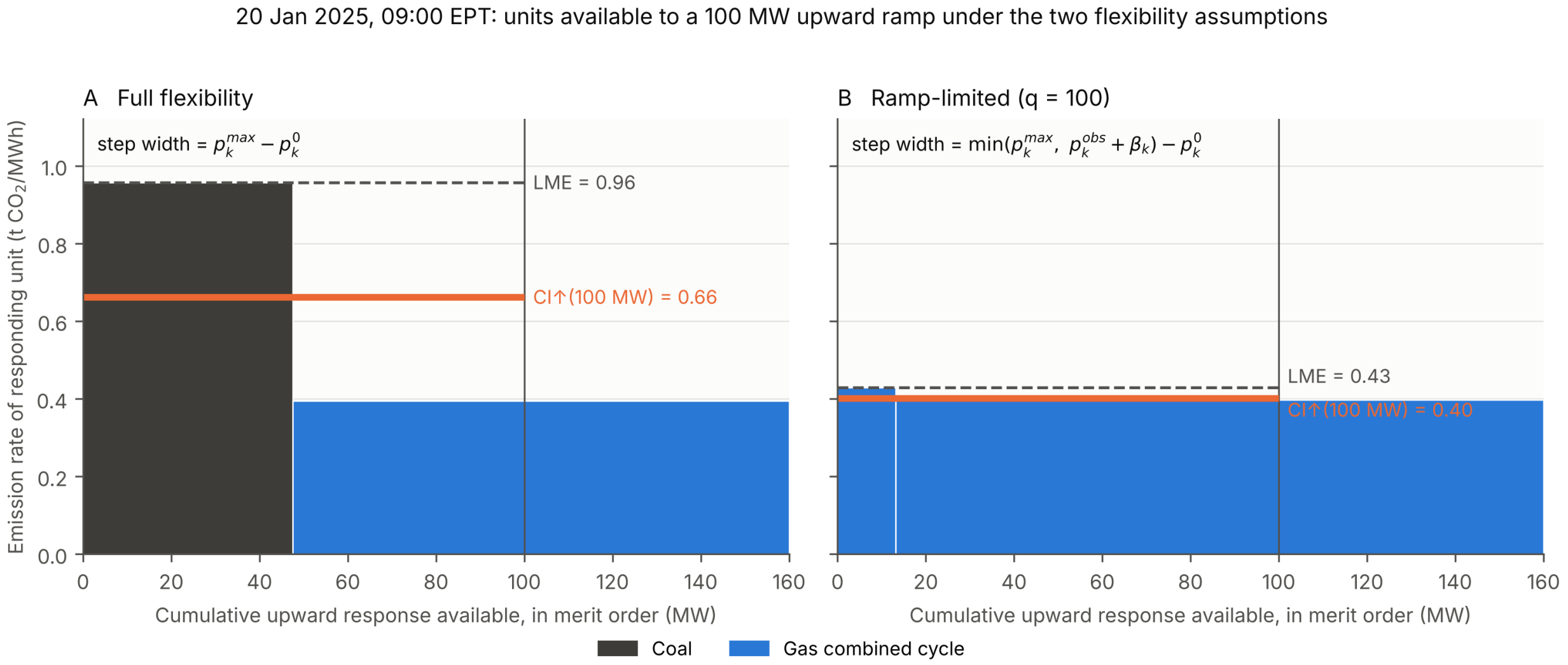

**Figure S3. The same hour under the two flexibility assumptions, related to Figure 1 and Experimental Procedures.** Units available to a 100 MW upward ramp at 20 January 2025, 09:00 EPT (the hour of Figure 1C), in merit order. Each step is one unit's upward headroom; its height is the unit's emission rate; the shaded area over 0 to 100 MW is the emission the ramp causes and its slope is $CI^{\uparrow}(100\,\text{MW})$. (A) Full flexibility: the headroom of unit $k$ is $p_k^{max} - p_k^0$. (B) Ramp-limited case ($q = 100$): each unit is anchored at its observed output within $\pm\beta_k$, which changes the base dispatch as well as the step widths, so the marginal unit differs between the panels (coal in A, gas combined cycle in B). Across January the marginal rate falls in the same group (gas, intermediate or coal, as in Figure 5) under the two assumptions in 59.95% of hours. Source: results/figS3_flexibility_steps.csv (real/figS3_flexibility.py).

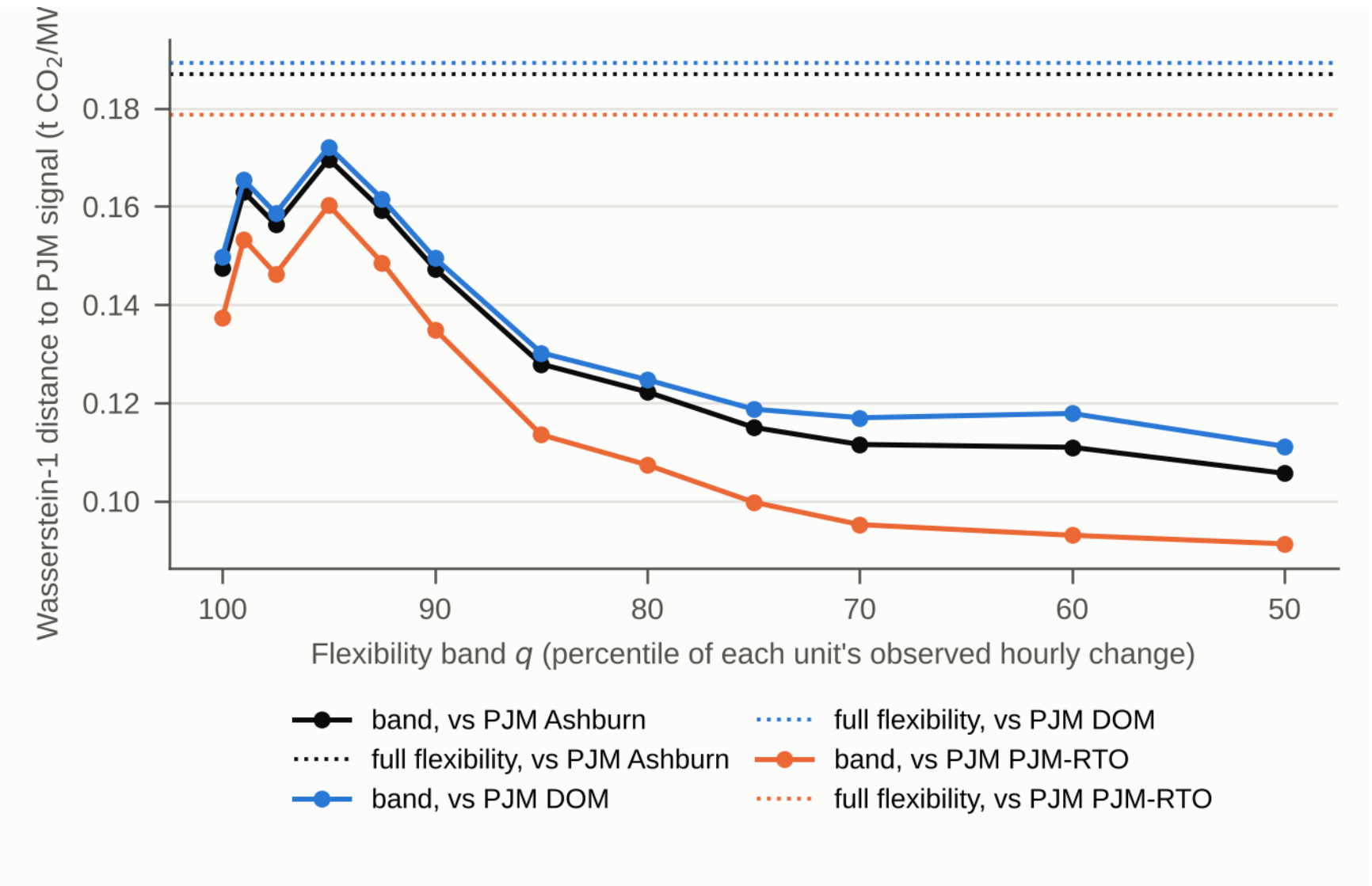

**Figure S4. The PJM five-minute signal does not identify a flexibility band, related to Figure 5 and Experimental Procedures.** Wasserstein-1 distance between the model's hourly upward LME for January 2025 and PJM's five-minute marginal emission rate at the Ashburn node, the Dominion zone and the PJM region, as a function of the band percentile $q$ (dotted lines: full flexibility). Both distributions are clipped to [0, 1.5] t $CO_2$/MWh. Against Ashburn the distance falls from 0.148 t $CO_2$/MWh at $q = 100$ to 0.106 at $q = 50$ with no minimum inside the range, and stays above 0.1 t $CO_2$/MWh for every band, because the model produces no negative and few intermediate marginal rates. Source: results/band_calibration.csv (real/calibrate_band.py).

## Supplemental tables

**Table S1. Sensitivity of the LME error share to the flexibility band, related to Figure 2 and Experimental Procedures.** Share of hours (%) in which the LME misstates an upward ramp by more than 10%, by ramp size, for the full-flexibility case and for bands set to the $q$th percentile of each unit's hourly changes; last column, the accounting gap $g(100\,\text{MW})$ (%). n.a.: in some hours the committed fleet cannot meet the campus plus ramp within the band, so the band is rejected as unphysical (Experimental Procedures). The ramp-limited case of the main text is $q = 100$. July was run for full flexibility and $q = 100$ only.

| Month | Case | 50 MW | 100 MW | 200 MW | 300 MW | 500 MW | 1,000 MW | $g(100\,\text{MW})$ |
|---|---|---|---|---|---|---|---|---|
| January | full flexibility | 11.02 | 20.30 | 36.83 | 50.00 | 66.94 | 83.20 | 46.58 |
| January | q = 100 | 12.50 | 24.06 | 46.51 | 59.54 | 75.13 | 87.50 | 42.52 |
| January | q = 99 | 17.07 | 29.97 | 48.92 | 64.25 | 80.91 | 90.46 | 43.02 |
| January | q = 97.5 | 17.61 | 30.78 | 56.45 | 69.62 | 83.87 | 92.20 | 42.81 |
| January | q = 95 | 17.74 | 35.08 | 57.53 | 74.73 | 85.08 | 91.53 | 42.70 |
| January | q = 92.5 | 23.52 | 45.30 | 67.88 | 76.21 | 86.42 | 92.34 | 42.33 |
| January | q = 90 | 24.60 | 45.30 | 65.86 | 77.55 | 85.89 | 93.55 | 41.14 |
| January | q = 85 | 31.32 | 48.66 | 67.61 | 77.96 | 86.56 | 93.95 | 38.29 |
| January | q = 80 | 36.42 | 56.85 | 74.19 | 84.14 | 92.61 | 95.16 | 36.89 |
| January | q = 75 | 40.19 | 60.08 | 74.19 | 87.23 | 92.20 | 96.37 | 35.55 |
| January | q = 70 | 40.32 | 59.01 | 71.37 | n.a. | n.a. | n.a. | 34.46 |
| January | q = 60 | 42.34 | 60.62 | 75.94 | n.a. | n.a. | n.a. | 33.55 |
| January | q = 50 | 71.91 | 85.75 | 93.41 | n.a. | n.a. | n.a. | 37.76 |
| July | full flexibility | 8.87 | 16.94 | 24.19 | 28.49 | 35.62 | 41.80 | 47.63 |
| July | q = 100 | 8.74 | 18.01 | 25.27 | 34.68 | 39.78 | 46.91 | 44.37 |

Source: results/band_sensitivity.csv and results/band_sensitivity_jul.csv (real/band_sensitivity.py).

**Table S2. Upward and downward metrics, full flexibility and ramp-limited cases, January and July 2025, related to Figures 2 and 4 and Note S1.** Share of hours in which the one-sided LME misstates the ramp by more than 10%, and accounting gap $g$, for upward and downward ramps of 100 MW and 1 GW.

| Month | Case | Ramp (MW) | Hours with $\eta^{\uparrow} > 10\%$ | Hours with $\eta^{\downarrow} > 10\%$ | $g^{\uparrow}$ | $g^{\downarrow}$ |
|---|---|---|---|---|---|---|
| Jan | full flexibility | 100 | 20.30% | 20.83% | 46.58% | 46.54% |
| Jan | full flexibility | 1,000 | 83.20% | 84.54% | 45.96% | 46.02% |
| Jan | ramp-limited | 100 | 24.06% | 22.85% | 42.52% | 42.12% |
| Jan | ramp-limited | 1,000 | 87.50% | 85.22% | 43.45% | 42.89% |
| Jul | full flexibility | 100 | 16.94% | 16.67% | 47.63% | 47.88% |
| Jul | full flexibility | 1,000 | 41.80% | 47.45% | 47.59% | 47.38% |
| Jul | ramp-limited | 100 | 18.01% | 16.13% | 44.37% | 44.11% |
| Jul | ramp-limited | 1,000 | 46.91% | 50.94% | 45.27% | 44.12% |

Source: results/downward_metrics.csv (real/downward_metrics.py; all six ramp sizes). The downward error share differs from the upward share by at most 1.88% of hours at 100 MW and 5.65% at 1 GW; the downward accounting gap differs from the upward gap by at most 1.15% of the caused emission. Direction matters for the value of the intensity in a given hour (Figure 3), not for how often the LME fails or for the size of the accounting gap.

**Table S3. Measured emission rates and heat rates by unit class, PJM balancing authority, January and July 2025, related to Figure 3 and Note S2.** Units with more than 100 MWh of gross generation in the month. $e$ in t $CO_2$/MWh and heat rate in MMBtu/MWh, both from the month's totals in the EPA CAMPD hourly data [S1]; p10 and p90 are the 10th and 90th percentiles across units.

| Month | Class | Units | $e$ p10 | $e$ median | $e$ p90 | Heat rate p10 | Heat rate median | Heat rate p90 |
|---|---|---|---|---|---|---|---|---|
| January | Coal | 68 | 0.82 | 0.90 | 1.03 | 8.8 | 9.6 | 10.8 |
| January | Gas combined cycle | 191 | 0.35 | 0.38 | 0.59 | 6.5 | 7.1 | 10.9 |
| January | Gas combustion turbine | 255 | 0.47 | 0.57 | 0.83 | 8.8 | 10.5 | 12.6 |
| January | Gas steam | 19 | 0.51 | 0.61 | 0.91 | 9.5 | 10.9 | 12.8 |
| January | Oil | 31 | 0.03 | 0.69 | 1.16 | 0.5 | 11.6 | 15.9 |
| July | Coal | 67 | 0.82 | 0.94 | 1.12 | 9.1 | 10.1 | 11.7 |
| July | Gas combined cycle | 190 | 0.35 | 0.38 | 0.60 | 6.5 | 7.1 | 11.5 |
| July | Gas combustion turbine | 286 | 0.48 | 0.57 | 0.73 | 9.2 | 10.9 | 13.4 |
| July | Gas steam | 21 | 0.52 | 0.57 | 0.79 | 9.7 | 10.6 | 12.9 |
| July | Oil | 42 | 0.06 | 0.75 | 1.24 | 9.1 | 12.1 | 18.9 |

Source: results/tableS3_emission_rates.csv. The oil class is heterogeneous (some units report heat input but little $CO_2$, giving rates close to zero) and supplies less than 1% of the response to any ramp (Figure 2B).

**Table S4. Sensitivity to the cost parameters, January 2025, full flexibility, related to Figure 2 and Experimental Procedures.** Share of hours (%) with LME error above 10% at 100 MW and 1 GW, share of hours with directional difference above 10% at 100 MW, and accounting gap $g(100\,\text{MW})$ (%), for the base cost parameters and seven alternatives. The daily gas price is recalibrated to observed coal output in every case, which is why the oil price has no effect (no oil unit is marginal in January) and why the other changes move any entry by at most 3.09.

| Case | $\eta > 10\%$, 100 MW | $\eta > 10\%$, 1 GW | $\delta > 10\%$, 100 MW | $g(100\,\text{MW})$ |
|---|---|---|---|---|
| Base (PJM VOM defaults, EIA fuel prices) | 20.30 | 83.20 | 37.63 | 46.58 |
| All VOM halved | 19.89 | 83.20 | 35.35 | 45.46 |
| All VOM raised by 50% | 19.09 | 82.80 | 34.54 | 47.87 |
| Flat VOM of 3 $/MWh | 21.64 | 81.85 | 40.19 | 45.33 |
| Coal 2.0 $/MMBtu | 20.03 | 83.33 | 35.22 | 47.16 |
| Coal 3.0 $/MMBtu | 18.68 | 83.87 | 36.29 | 45.68 |
| Oil 10 $/MMBtu | 20.30 | 83.20 | 37.63 | 46.58 |
| Oil 25 $/MMBtu | 20.30 | 83.20 | 37.63 | 46.58 |

Source: results/cost_sensitivity.csv (real/cost_sensitivity.py).

# Supplemental notes

## Note S1. Metrics for downward ramps, related to Figures 2 to 4

The main text reports the error of the LME, the directional difference and the accounting gap for upward ramps. This note gives the downward counterparts and their values. Symbols are as in Experimental Procedures:

$E_h(\theta)$ is the emission of the system dispatch in hour $h$ when the campus draws $\theta$ (t/h), $\theta_0$ is the campus base load (MW), $\Delta > 0$ is the ramp size (MW), $\varepsilon = 0.1$ MW, and $a_h$ is the regional average intensity (t $CO_2$/MWh).

### S1.1 Downward directional intensity and one-sided LME

A downward ramp reduces the campus load from $\theta_0$ to $\theta_0 - \Delta$. The emission it avoids per MWh is

$$CI_h^{\downarrow}(\Delta) = \frac{E_h(\theta_0) - E_h(\theta_0 - \Delta)}{\Delta},$$

and the one-sided LME for a load decrease is

$$\lambda_h^{\downarrow} = \frac{E_h(\theta_0) - E_h(\theta_0 - \varepsilon)}{\varepsilon}.$$

Both are positive when reducing load reduces emissions. $CI_h^{\downarrow}(\Delta) = \lambda_h^{\downarrow}$ for every $\Delta$ within the downward validity range, the distance below $\theta_0$ over which the marginal emission rate is constant. The units that absorb a downward ramp are those at and below the margin in merit order, whereas an upward ramp is met by those at and above it, which is why $CI_h^{\downarrow}(\Delta)$ and $CI_h^{\uparrow}(\Delta)$ generally differ once $\Delta$ exceeds either validity range.

### S1.2 LME error for a downward ramp

$$\eta_h^{\downarrow}(\Delta) = \frac{\left|\lambda_h^{\downarrow} - CI_h^{\downarrow}(\Delta)\right|}{\left|CI_h^{\downarrow}(\Delta)\right|},$$

computed over hours with $|CI_h^{\downarrow}(\Delta)| \geq 0.05$ t $CO_2$/MWh. Table S2 reports the share of hours with $\eta_h^{\downarrow}(\Delta) > 0.1$ next to the upward share.

### S1.3 Accounting gap for a downward ramp

Under location-based reporting, a campus that reduces its load by $\Delta$ in hour $h$ reduces its declared emission by $a_h\Delta$. The emission actually avoided is $CI_h^{\downarrow}(\Delta)\Delta$. The difference,

$$\left(CI_h^{\downarrow}(\Delta) - a_h\right)\Delta,$$

is avoided but not credited to the campus; it lowers the regional average and is credited, in proportion to consumption, to all other consumers. Assuming the campus ramps down by $\Delta$ in every hour of the month, the share of the avoided emission that is not credited to the campus is

$$g^{\downarrow}(\Delta) = 1 - \frac{\sum_h a_h}{\sum_h CI_h^{\downarrow}(\Delta)},$$

with both sums over hours with $CI_h^{\downarrow}(\Delta) > 0.05$ t $CO_2$/MWh. A positive $g^{\downarrow}$ means that location-based reporting under-credits the campus for the emission its downward ramps avoid, by the same mechanism that under-charges it for upward ramps.

### S1.4 Directional difference

The directional difference $\delta_h(\Delta)$ in Experimental Procedures already uses both directions and needs no downward counterpart. It is the gap between $CI_h^{\uparrow}(\Delta)$ and $CI_h^{\downarrow}(\Delta)$ relative to their mean, so that a pair (0.3, 0.1) t $CO_2$/MWh gives $\delta = 1$ (a 100% difference) whichever direction is taken as reference.

### Note S2. Why the fleet's emission rates fall into three groups, related to Figure 3

The emission rate of a unit (t $CO_2$/MWh) is the product of the carbon content of its fuel (t $CO_2$ per MMBtu of heat input) and its heat rate (MMBtu of heat input per MWh of gross generation):

$$e_k = \phi_{f(k)}\, r_k,$$

where $\phi_{f(k)}$ is the carbon content of the fuel $f(k)$ of unit $k$ and $r_k$ is its heat rate. In the model both $e_k$ and $r_k$ are measured, as the ratio of the month's $CO_2$ mass to gross generation and of the month's heat input to gross generation, from the EPA CAMPD continuous emission monitoring data [S1]; $\phi$ is not used by the model and appears here only to explain the grouping. The default emission factors of the U.S. EPA Greenhouse Gas Reporting Program [S2] give a carbon content of 0.05306 t $CO_2$/MMBtu for natural gas and 0.09328 (bituminous) to 0.09717 (sub-bituminous) t $CO_2$/MMBtu for the coals burnt in PJM. Gas-fired classes therefore differ only in heat rate: at the January fleet medians (Table S3), a combined-cycle unit, which recovers heat from its exhaust to drive a steam turbine, burns 7.13 MMBtu/MWh, a simple-cycle combustion turbine 10.48 and a gas steam unit 10.90; coal units burn 9.60 MMBtu/MWh at nearly twice the carbon content. The products of these heat rates with the carbon contents are 0.38, 0.56, 0.58 and 0.90 to 0.93 t $CO_2$/MWh, against measured medians of 0.38, 0.57, 0.61 and 0.90 t $CO_2$/MWh; the largest difference is 0.03 t $CO_2$/MWh (gas steam), which reflects part-load operation and fuel blending in the measured data.

The three groups matter for the size effect. A ramp met within the gas combined-cycle group or within the coal group changes the marginal rate by at most the spread within that group (0.24 and 0.21 t $CO_2$/MWh between the 10th and 90th percentiles in January); a ramp that crosses from the combined-cycle group to the coal group changes it by the distance between their medians, 0.52 t $CO_2$/MWh. Combustion turbines and gas steam units sit between the two (medians 0.57 and 0.61 t $CO_2$/MWh) with wider spreads (0.36 and 0.40), so a ramp that passes through them climbs in smaller steps. The error of the LME for a finite ramp is therefore large in a fleet whose rates form well-separated groups and small in a fleet of one fuel, which is the mechanism behind the seasonal pattern in Figure 2: in July, combustion turbines and steam units in the middle group supply a larger share of the ramp, so fewer ramps cross directly from the gas combined-cycle group to the coal group (Figure S1).

## Supplemental references